\documentclass[preprint,pteplogo]{ptephy_v2}
\usepackage{hyperref}

\usepackage{amssymb}
\usepackage{amsmath}
\usepackage{textcomp}
\usepackage{mathcomp}
\usepackage{longtable}
\usepackage{lscape}
\usepackage{natbib}

\begin{document}




\title{Monte Carlo Simulations of Secondary Cosmic-Ray Variations in Atmospheric Electric Fields : Implications for Long Duration Electron and Gamma-ray Emissions from Thunderclouds}


\author{H. Tsuchiya} 
\affil{Nuclear Science and Engineering Center, Japan Atomic Energy Agency, Tokai-mura, Naka-gun,
            319-1195, 
        Ibaraki,Japan
        \email{tsuchiya.harufumi@jaea.go.jp}
        }

\begin{abstract}
Monte Carlo simulations were conducted using the Particle and Heavy Ion Transport code System (PHITS) to investigate the role of secondary cosmic rays in the generation of long-duration bursts from thunderclouds and to clarify the conditions of the electric field region responsible for particle acceleration. The simulations utilized realistic secondary cosmic-ray spectra, including gamma rays, electrons, positrons, and muons, as input. The simulation results indicate that gamma rays provide the dominant supply of seed electrons for long-duration bursts, regardless of the geometry or strength of the electric field region. They also reveal the structure and strength of the electric field region required to produce gamma rays exceeding several tens of MeV, which 
have so far been detected only by high-altitude observations. Furthermore, the fluxes of long-duration bursts estimated from the simulation results were compared with observational data to constrain the properties of the electric field region. In particular, the comparison with measurements at Yangbajing, located at an altitude of 4.3~km, helps 
narrow down the possible range of electric field strengths and configurations.
\end{abstract}
\subjectindex{A22, F04, J63}

\maketitle




\newcommand{\ef}{\mathrm{kV\,m^{-1}}}
\section{Introduction}
\label{sec:intro}
High-energy radiation bursts emitted from thunderclouds have been observed for durations ranging from a few seconds to several tens of minutes by various detectors including 
ground-based~\cite{torii_observation_2002,tsuchiya_detection_2007,kuroda_observation_2016,wada_termination_2018,wada_gamma-ray_2019,Tsurumi_GRL_2023}, 
high-altitude~\cite{EAS-top_2000,chubenko_intensive_2000,tsuchiya_observation_2009,tsuchiya_observation_2012,gurevich_observations_2016,chilingarian_catalog_2019}, and airborne instruments~\cite{mccarthy_further_1985,eack_xray_1996,kelley_relativistic_2015,ostgaard_glow_20km_2019}. 
These long-duration bursts, often referred to as "gamma-ray glow" or "terrestrial ground enhancement",
are thought to result from electrons accelerated to relativistic energies by the 
electric fields (EF) in thunderclouds.
In fact, sea-level observations along the coast of the Japan Sea have recorded gamma rays
with energies as high as 20~MeV~\cite{wada_catalog_2021}, 
while high-altitude observations have detected 
gamma rays exceeding 40~MeV~\cite{tsuchiya_observation_2012,chilingarian_thunderstorm_spectrum_2013}. In addition to gamma rays, count increases due to electrons have been also reported~\cite{muraki_effects_2004,chilingarian_particle_2011}.
Although these emissions are generally believed to be attributed to electrons accelerated in 
EF regions, a critical question remains: How can electrons
be accelerated to relativistic energies by an EF while overcoming ionization losses in the atmosphere? 

This question primarily arises from the EF strength that is actually measured in the atmosphere.  Observed EF strength in the atmosphere is at most about one-tenth or less of the magnitude required to cause conventional atmospheric electrical breakdown~\cite{marshall_observed_1995,MacGorman1998TheEN,marshall_observed_2005,stolzenburg_duration_2010,atmos12121645}. As pointed out by {\it e.g.}, \citet{gurevich_runaway_1992} and \citet{dwyer_implications_2004}, such EFs would decelerate electrons with energies below approximately 0.1~MeV, since the ionization losses exceed the energy gained from the EF at these energies. Therefore, seed electrons with energies of approximately 0.1~MeV or higher are considered necessary to reach relativistic energies under the observed EF strengths. 
Several previous studies of long-duration bursts have emphasized that the electron component of
secondary cosmic rays plays a central role in EF-induced particle enhancement
(e.g., \cite{chilingarian_atmospheric_2015,cramer_simulation_2017}).
Other studies have considered various components of secondary cosmic rays,
including electrons and additional particle species, as possible sources of
seed electrons (e.g., \cite{kelley_relativistic_2015,ostgaard_glow_20km_2019}).
However, it remains unclear which component most effectively supplies
high-energy electrons inside thundercloud electric fields, and a systematic
evaluation of the relative contributions from different secondary cosmic-ray
species has not yet been performed. Such knowledge is also essential for
realistic modeling of electron acceleration under atmospheric electric-field conditions.

High-energy seed electrons can be accelerated by realistic electric fields, leading to 
ionization of air molecules and the emission of bremsstrahlung gamma rays. This process initiates a chain reaction in which newly generated high-energy electrons are further accelerated, causing additional ionization and photon production. This cascade process is known as Relativistic Runaway Electron Avalanches (RREAs) and is considered a fundamental mechanism underlying 
long-duration bursts. Simulation studies~\cite{dwyer_fundamental_2003,babich_fundamental_2004} have predicted that an EF strength of
284~$\mathrm{kV\,m^{-1}}$ at standard temperature and pressure (STP; sea level) is required to initiate RREA. This threshold is proportional to atmospheric density and thus decreases with increasing altitude. 
At a given altitude of $z$ km, the threshold is expressed as 
$284\times n/n_0$~$\ef$, 
where $n$ is the atmospheric density at altitude $z$ km, and $n_0$ is the atmospheric density at STP. 
In addition to the RREA process, the concept of a "Modification Of Spectrum" has been proposed~\cite{Chilingarian_MOS_2014}. It suggests that even when the EF strength is below the 
RREA threshold, the electron component of secondary cosmic rays, particularly
high-energy electrons ($>$1~MeV) can still be accelerated and
produce long-duration bursts without triggering a full RREA. 

High-altitude observations of long-duration bursts have detected gamma rays with 
energies $>$40~MeV~\cite{tsuchiya_observation_2012,chilingarian_thunderstorm_spectrum_2013}. While these gamma rays are generally attributed to the acceleration of electrons, it remains unclear 
whether secondary cosmic-ray electrons alone are sufficient to account for 
the source of such high-energy gamma rays. 
In fact, secondary cosmic-ray gamma rays and muons 
can also generate electrons via electromagnetic interactions. Furthermore, 
positrons and positive muons, both of which are accelerated in the opposite direction to 
electrons in an EF,  may additionally supply high-energy electrons in the EF region, 
potentially contributing to the generation of long-duration bursts.

The present study focuses on how steady-state secondary cosmic rays, rather than transient particles related to extensive air showers, are modified while traversing localized thundercloud electric fields, and how this modification affects the particle composition and fluxes reaching the ground. This perspective is important since high-altitude observations, such as those at Yangbajing~\cite{tsuchiya_observation_2012,tsuchiya_tibet_2024} and 
Mt.~Aragatz~\cite{chilingarian_catalog_2019}, have detected gamma rays exceeding 40~MeV. However, the electric-field structures and acceleration conditions capable of producing such high-energy 
gamma rays in thunderclouds are still not well understood.
By systematically treating multiple particle species and EF geometries, the present work aims to identify which components of secondary cosmic rays effectively supply high-energy electrons inside the EF region and to evaluate whether such an electron supply can account for the magnitude level of the observed long-duration bursts.

To address this aim, we perform Monte Carlo (MC) simulations to 
examine the behaviors of various particles, including gamma rays, 
electrons, positrons, and muons, within an EF region. 
The resulting electron and gamma-ray spectra, 
together with their spatial distributions, are calculated.  
Based on these calculations, we discuss the extent to which secondary cosmic rays 
contribute to long-duration bursts and the electric field structure capable of
generating gamma rays with energies exceeding 40~MeV.

\section{Methods}
\label{sec:methods}
\begin{figure}[tbh]
\centering
\includegraphics[scale=0.30]{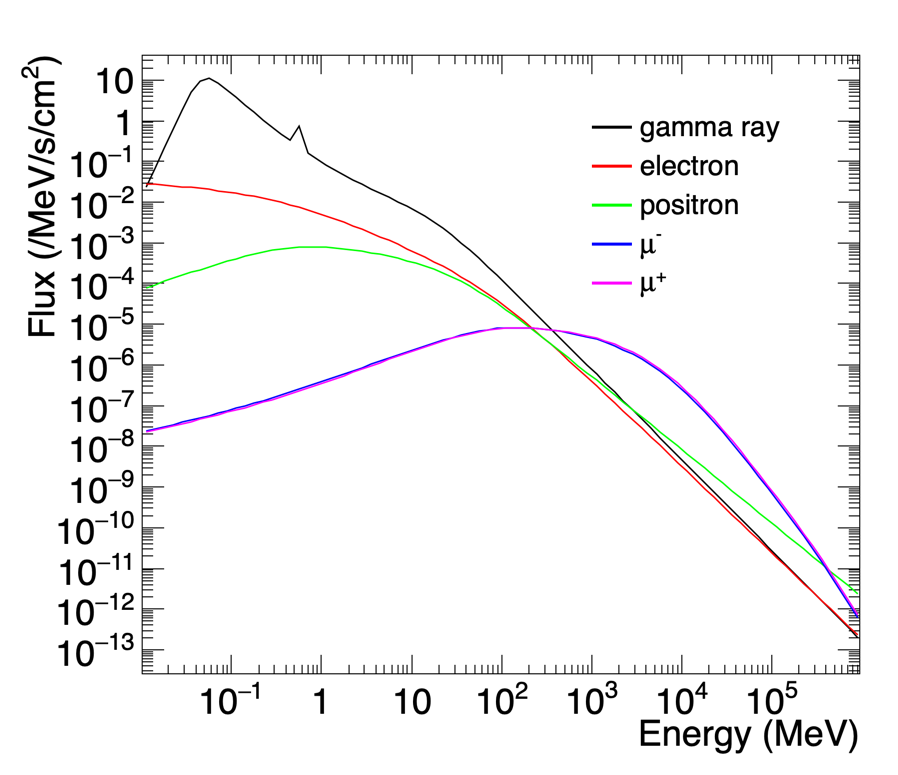}
\caption{Secondary cosmic-ray spectra expected at Yangbajing (4.3 km a.s.l.), calculated by EXPACS~\cite{EXPACS} }\label{fig:2ndcrspe}
\end{figure}
To evaluate the EF effects on secondary cosmic rays at an altitude of 4.3~km above sea level (a.s.l.), 
we employed the Particle and Heavy Ion Transport code 
System (PHITS ver.3.34)~\cite{PHITS}.
PHITS natively incorporates the EXPACS model~\cite{EXPACS}, which 
generates secondary cosmic-ray spectra as a function of altitude, latitude and longitude.
In recent studies, EXPACS-generated spectra 
 have been often utilized as input for GEANT4 and other simulation frameworks to investigate particle development in atmospheric electric fields~\cite{cramer_simulation_2017,bowers_combining_2019,Diniz_2023_GEANT4,sarria_2023_GEANT4}. Therefore, using PHITS directly avoids the need for external coupling and provides a consistent and streamlined framework for this study.

Figure~\ref{fig:2ndcrspe} presents various secondary cosmic-ray spectra utilized in this study, corresponding to those expected at Yangbajing, Tibet, at 4.3~km a.s.l. (90.522$\tcdegree$E and 30.102$\tcdegree$N). 
These spectra, including gamma rays, electrons, positrons, and negative and positive muons 
(denoted as, $\mu^-$ and $\mu^+$, respectively), serve as the initial inputs for our 
simulations. 
To validate the secondary cosmic rays used in this simulation, 
we estimated the expected count rate of the Yangbajing neutron monitor 
based on these spectra. The resulting count rate was found to 
agree with the observed value within approximately 30\% (see Supplement information 1).

PHITS incorporates important physical processes relevant to the production of long-duration bursts. The interactions of gamma rays with electrons (and positrons) include the photoelectric effect,  
Compton scattering, and pair production. For electrons, positrons, and muons, fundamental processes  such as bremsstrahlung, ionization losses, Coulomb scattering, and elastic scattering are taken into account. Although neutron propagation in the atmosphere is not explicitly analyzed 
in this work, neutron production via photonuclear reactions is considered in the simulations, 
with nuclear data given by the JENDL-4 library~\cite{jendl4}.

\begin{figure}[tbh]
\centering
\includegraphics[scale=0.20]{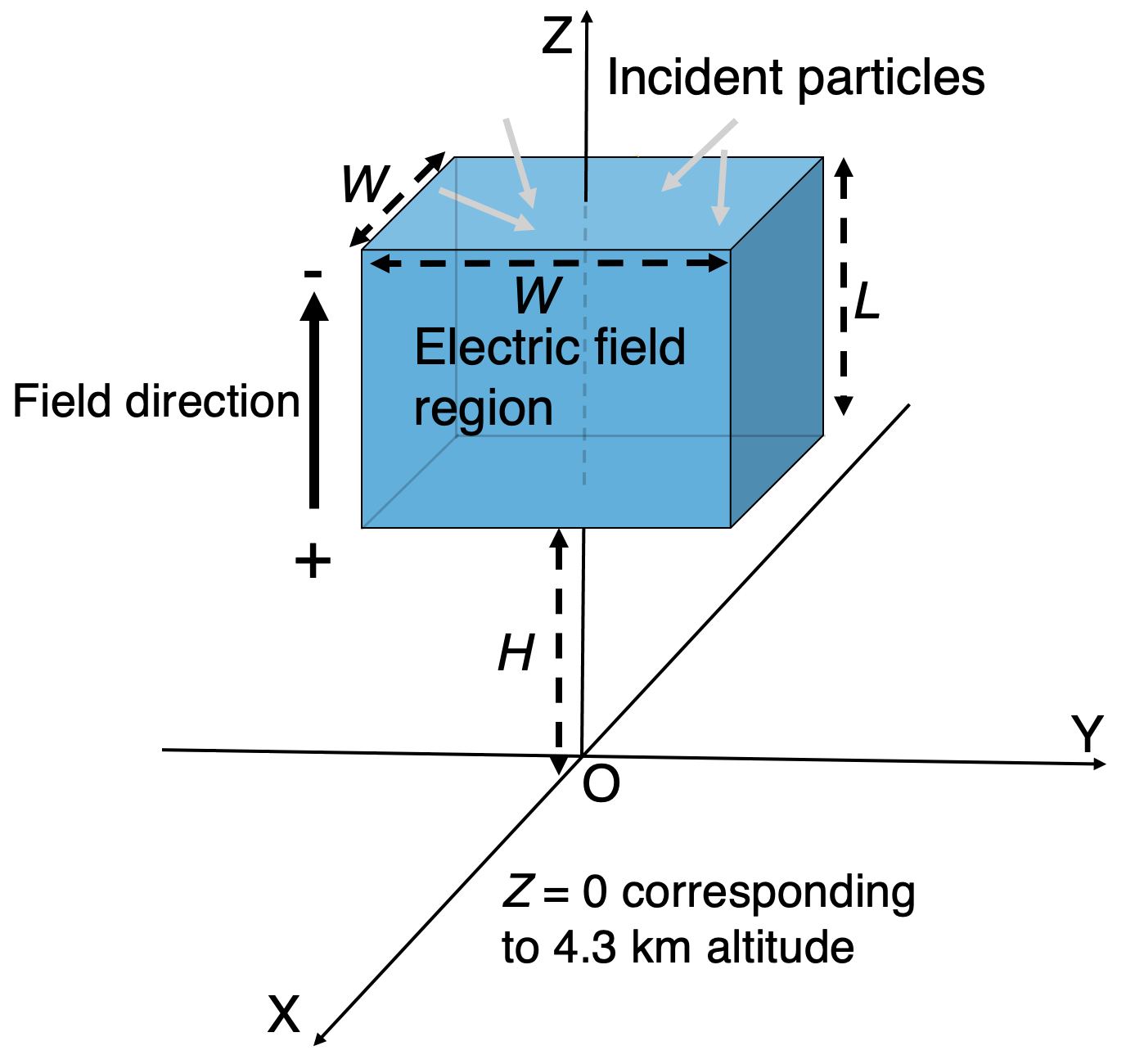}
\caption{Assumed structure and orientation of the electric field. The assumed EF region is shown in light blue. Dashed arrows indicate the parameters $W$, $L$, and $H$ used in the simulations. The EF orientation is represented by a solid arrow. The origin of the simulation coordinate system is denoted as O.
Particles are injected from the upper boundary of the electric field region.}\label{fig:EF_structure}
\end{figure}

The minimum incident energy was set to 50 keV for all particle types. 
For reference, 
the integrated flux (cm$^{-2}$s$^{-1}$) above 50~keV for each particle shown in Fig.~\ref{fig:2ndcrspe} 
is 1.2 for gamma rays, $3.6\times10^{-2}$ for electrons, $1.7\times10^{-2}$ for positrons, 
$1.5\times10^{-2}$ for $\mu^{-}$, and $1.7\times10^{-2}$ for $\mu^{+}$.
Each particle, following its energy spectrum (Fig.~\ref{fig:2ndcrspe}), was injected 
just above the EF region depicted in Figure~\ref{fig:EF_structure}.  
The dependence of the zenith angle, $\theta$, for each injected particle is presented in the supplementary materials and is represented by the expression $I_\mathrm{0}\cos^{n}(\theta)$, 
where $I_\mathrm{0}$ denotes the vertical intensity. 
The range of $\theta$ was assumed to be from 0$\tcdegree$ to $60\tcdegree$.
The power index $\mathrm{n}$ varies for each particle, and typically ranges from 2 to 3.5 (see the Supplementary information 2). 
Detailed information on the angular distributions implemented in PHITS can be found in~\cite{sato_angdist_PHITS}.

The atmospheric composition in the simulations was assumed to be $75.5$\% $^{14}\mathrm{N}$, $23.2$\% $^{16}\mathrm{O}$, and $1.3$\% $^{40}\mathrm{Ar}$. 
The altitude of the ground surface was set at 4.3~km, and the atmospheric density was
derived from the U.S. Standard Atmosphere Model as 
\begin{equation*}
  1.218\times10^{-3}\exp(-z/9.9\, \mathrm{km})\,\, \mathrm{g\, cm}^{-3}, 
\end{equation*}
where $z$ denotes the altitude above sea level. The density was 
updated every 0.5~km along the vertical direction. 
Since the ground itself was not modeled in this study, particles reaching the 4.3~km altitude
were no longer tracked.

As shown in Figure~\ref{fig:EF_structure}, the assumed EF region is modeled as a cuboid with a horizontal width of $W$ and a vertical length of $L$. The horizontal direction 
of an EF region is parallel to the X- or Y-axis direction, while the vertical direction is 
along the Z-axis direction.
The values of $W$ and $L$ are changed according to simulations, ranging from 100~m to 1250~m. 
In the present study, we investigate how these two parameters, $W$ and $L$, influence the production and transport of electrons and gamma rays in the EF region. In particular, the ratio $W/L$ characterizes the importance of lateral particle escape due to scattering, thereby affecting the efficiency with which electrons and gamma rays propagate through the EF region and contribute to the particle flux at its bottom.

We also introduce an additional spatial parameter, $H$ (Fig.~\ref{fig:EF_structure}), defined as the distance from the bottom of the EF region to the ground surface ($Z = 0$ in Fig.~\ref{fig:EF_structure}). The parameter $H$, corresponding to the cloud-base height, governs the survival probability of electrons
and gamma rays reaching the ground after being emitted from the EF region.

The EF is assumed to be homogeneous within the defined region, and its orientation is set such that it accelerates electrons toward the ground surface.
The EF strength used in the simulations ranges from 0 to 260~$\ef$.
An EF strength of 170~$\ef$ is equivalent to the value of 
284~$\ef$ at sea level required to induce RREA in the standard atmosphere~\cite{dwyer_relativistic_2007}, scaled to the altitude of 4.3~km. 
Consequently, the maximum EF strength considered in this study, 
260~$\ef$ is about 1.5 times the RREA threshold at this altitude.
In terms of the EF direction, this study focuses on the acceleration direction of negatively charged particles, such as electrons, toward the ground (Fig.~\ref{fig:EF_structure}).

To investigate how EF affects each secondary cosmic-ray particle,
$10^5$ particles were randomly injected over a $W \times W$ area on the upper 
surface of the EF region.
Subsequently, the energy spectra of electrons and gamma rays
escaping from the EF region were 
obtained both at the ground surface and just below the bottom of the EF region. 
In these samplings, particles located within 2~km from the central axis of the 
EF region (i.e., the $Z$ axis) were included and the resulting fluxes were normalized 
by the corresponding area.
In addition, the spatial distribution of the secondary particles generated by the 
incident particles was analyzed to explore its dependence on the strength and structure of the EF.

As mentioned above, secondary cosmic-ray particles are injected through the upper surface of the EF region,
while particles entering through the lateral boundaries are not explicitly considered. 
This treatment, commonly adopted in previous studies~\cite{Zhou_2016_AROG_YBJ, sarria_2023_GEANT4}, 
enables us to focus on the acceleration and multiplication processes inside the EF region. 
The secondary cosmic-ray flux is dominated by downward-moving particles, and the EF is oriented vertically. 
In addition, the effectiveness of particle acceleration depends on the effective path length of particles along the vertical extent $L$ 
of the EF region. This implies that particles entering from the side boundaries generally traverse shorter path lengths inside the EF, 
unless they have highly inclined trajectories. Consequently, the main contribution to
EF-induced enhancement of electrons and gamma rays is expected to originate 
from particles entering close to the upper boundary. 
We note that for small values of 
$W$ compared to $L$, contributions from lateral injection may become non-negligible,
as the effective area of the side boundaries ($W \times L$) increases relative to that of the upper surface ($W\times W$). A more realistic treatment including lateral injection
with altitude-dependent spectra will be addressed in future work.

\section{Results}
\subsection{Spatial distributions}
\begin{figure}[tbh]
\centering
\includegraphics[scale=0.40]{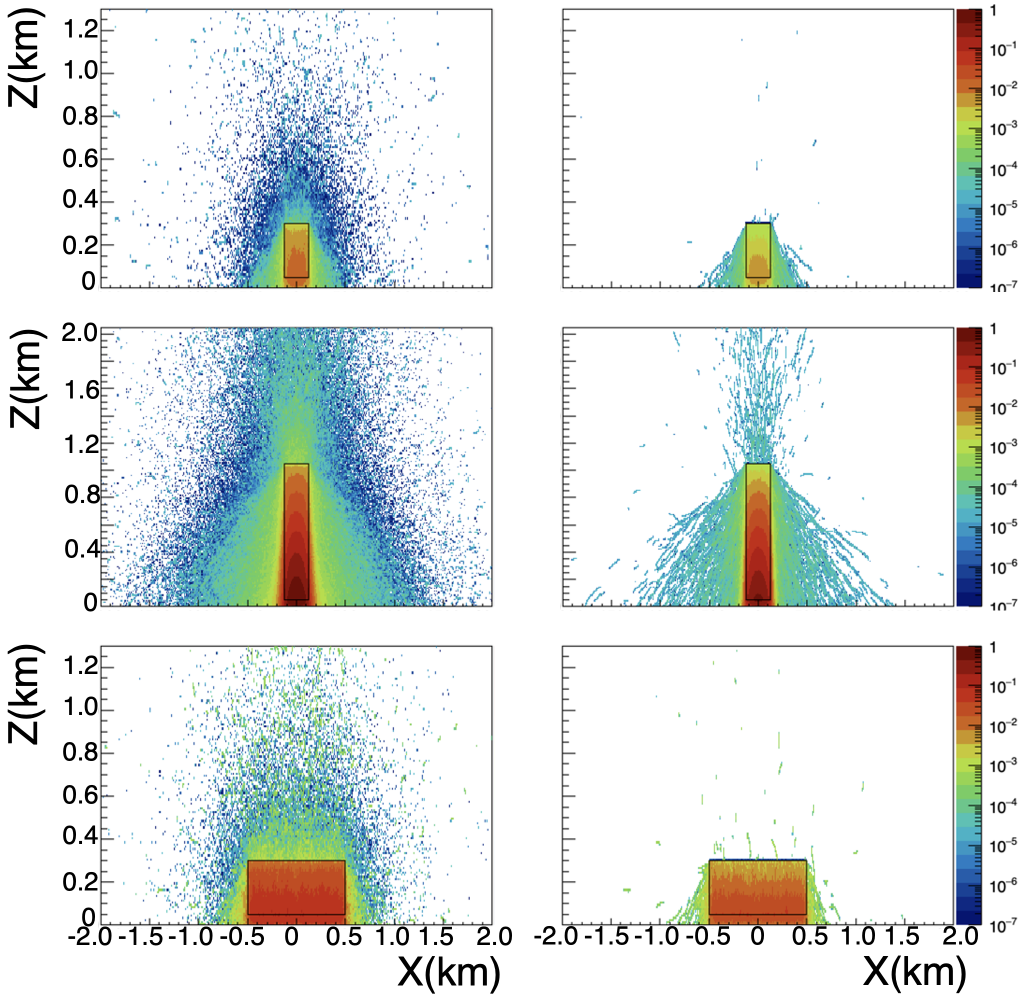}
\caption{
Spatial distributions of electron flux (cm$^{-2}$~src$^{-1}$) produced by various incident particles. The top, middle, and bottom panels correspond to electrons originating from $(W, \, L)=(250\, \mathrm{m}, 250\, \mathrm{m})$, $(W,\, L)=(250\, \mathrm{m}, 1000\, \mathrm{m})$, and $(W,\, L)=(1000\, \mathrm{m}, 250\, \mathrm{m})$,
respectively. The rectangles in all panels indicate the 
EF regions assumed in the MC simulations. The EF strength is 260~$\ef$.
The left panels show electrons with energies between 0.05 and 10~MeV, 
while the right panels show electrons with energies above 10~MeV. 
The horizontal and vertical axes represent the spacial extent in meters. 
Warmer colors (red) indicate higher fluxes.
}\label{fig:electron_L250_1000_E2.6}
\end{figure}

\begin{figure}[tbh]
\centering
\includegraphics[scale=0.40]{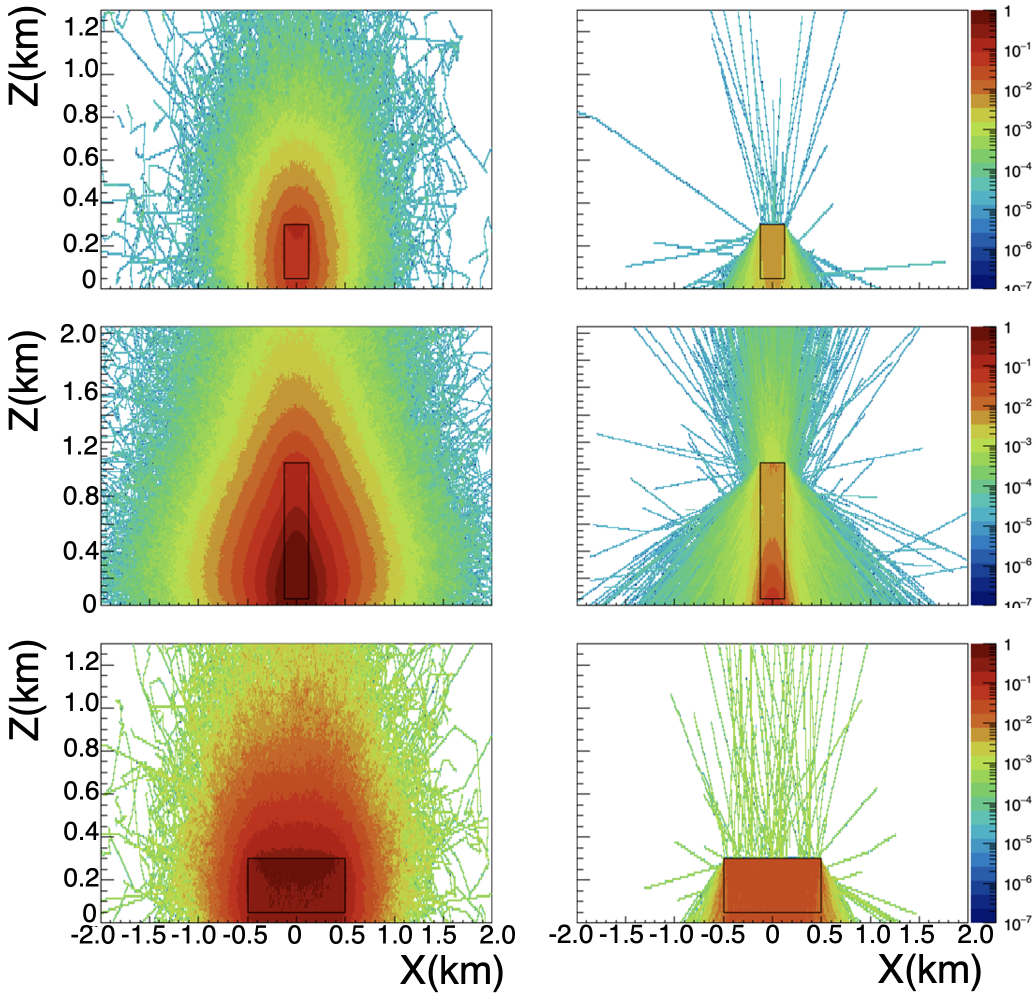}
\caption{The sames as Fig.~\ref{fig:electron_L250_1000_E2.6}, but for gamma rays 
in the 0.01-10~MeV (left) 
and $>$10~MeV (right). 
}\label{fig:photon_L250_1000_E2.6}
\end{figure}

Figures~\ref{fig:electron_L250_1000_E2.6} and \ref{fig:photon_L250_1000_E2.6} show examples of the spatial distributions 
of electrons and gamma rays, respectively, for several combinations of the EF-region parameters $L$ and $W$, with a fixed EF strength 
of 260~$\ef$ (In Supplementary Information, Figure S3 provides spatial distributions of electron flux under the EF strength of 170~$\ef$). 
To obtain these results,  all relevant components of secondary cosmic rays were injected into the EF region using the spectra 
shown in Fig.~\ref{fig:2ndcrspe}. The relative contributions of each particle were considered by weighting their fluxes accordingly. 
This process was conducted utilizing the multisources mode of PHITS, which automatically handles the injection of multi particle
components with prescribed spectral shapes and relative intensities.
The EF region is assumed to have several dimensions shown in the figures as rectangles, with a cloud base height ($H$) of 50~m.
In the two Figures, a total of $10^5$ source histories were generated in the multisource mode, 
and the resulting spatial fluxes are normalized per source history (cm$^{-2}$ src$^{-1}$). Therefore, 
the figures are intended to illustrate the normalized spatial distributions rather than fluxes integrated over a common time period.

Fig.~\ref{fig:electron_L250_1000_E2.6} clearly indicates that electron population increases 
near the bottom of the EF region.
These electrons are primarily produced via Compton scattering, pair production, and ionization,
and are subsequently accelerated toward the ground.
Figure~\ref{fig:photon_L250_1000_E2.6} illustrates spatial distributions of gamma rays, which are broader than those of electrons due to their high penetration ability. 
The overall behavior of gamma rays is generally similar to 
that of electrons, as gamma rays are predominantly produced via electromagnetic
processes related to electrons accelerated in the EF region.
In the present study, the electric field is oriented such that 
electrons are accelerated downward toward the ground. As a consequence, the RREA 
develops progressively as electrons propagate through the EF region, and the 
electron population grows along the downstream direction of acceleration. This naturally 
results in the maximum fluence of electrons and gamma rays appearing near the bottom of 
the EF region.
Clear upward-moving components are also observed in gamma-ray and electron flluences.
These upward-moving components mainly originate from back-scattered Compton photons and upward-accelerated positrons. 
The spatial distribution of positrons corresponding to these upward components is shown in Supplementary Information as Figure~S4, which confirms that positrons are accelerated upward toward the upper part of the EF region. This behavior is consistent with expectations from the RREA theory.

\begin{figure}[bth]
\centering
\includegraphics[scale=0.3]{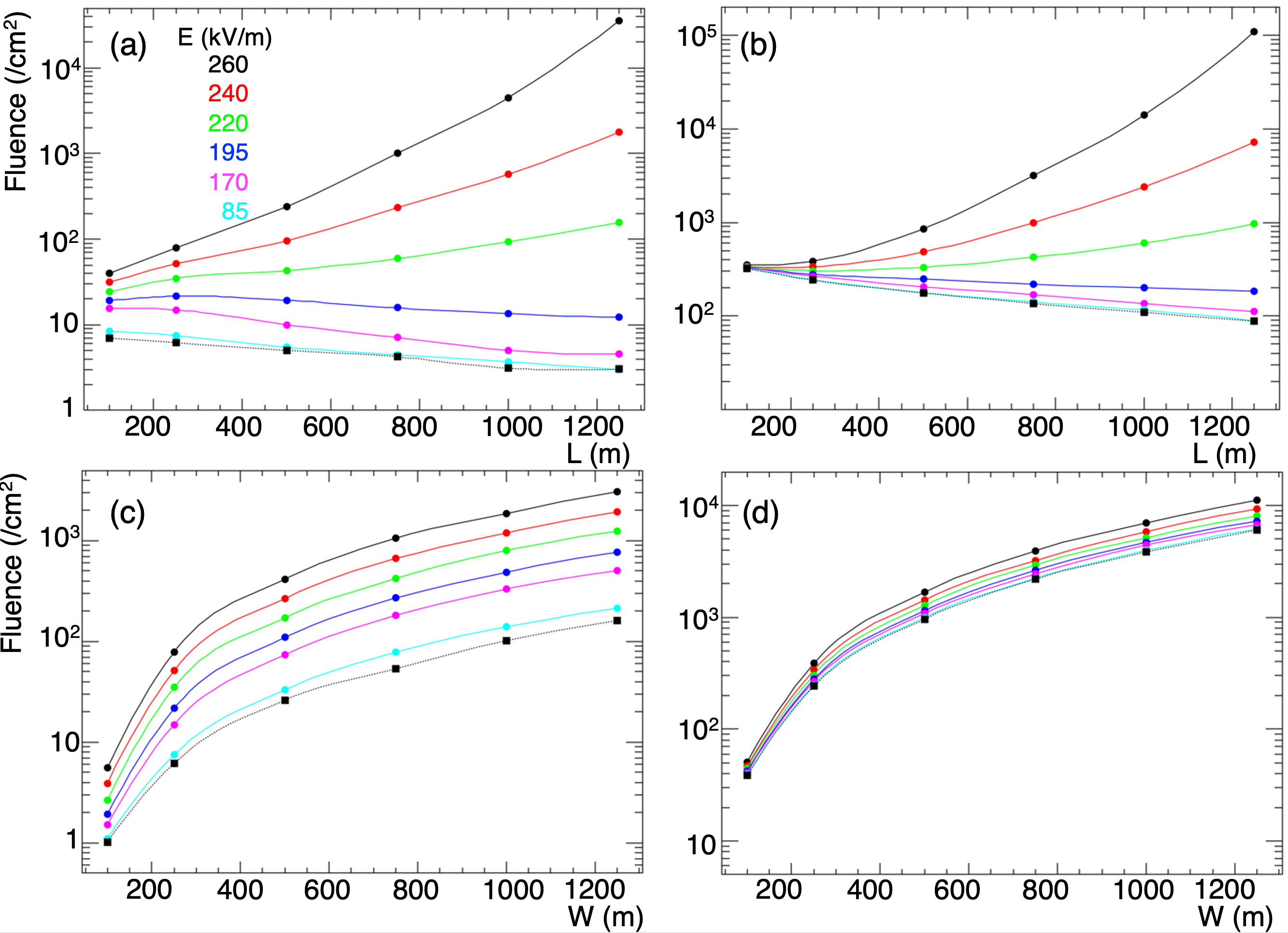}
\caption{Electron (left) and gamma rays (right) fluences (cm$^{-2}$) in 
0.05-10~MeV as a function of $L$ for panels (a) and (b), and as a function of $W$ 
for panels (c) and (d). Fluences in panels (a) and (b) 
were obtained with $W =$ 250~m, while those in (c) and (d) were obtained with $L=$ 250~m. 
Each colored circle corresponds to an EF strength ranging from 85 to 260~$\ef$. Black squares indicate electron fluences 
at 0~$\ef$, corresponding to the lowest values in each panel.
}\label{fig:electron_photon_L_W_E}
\end{figure}

EF regions with a longer $L$ or a wider $W$ tend to increase
the number of electrons and gamma rays (Figs.~\ref{fig:electron_L250_1000_E2.6} and \ref{fig:photon_L250_1000_E2.6}). 
To better understand these dependencies, 
Figure~\ref{fig:electron_photon_L_W_E} depicts the electron and gamma-ray fluence just below the EF region 
as a function of $L$ and $W$. The electron and gamma-ray fluence generally decrease with increasing $L$, 
unless the EF strength exceeds a certain threshold, which is approximately 195~$\ef$ in this
study [panels (a,b)]. In contrast, they increase with $W$, regardless of the EF strength [panels (c,d)]. 
However, the gamma-ray fluence increases with $W$, although its dependence on EF strength is weaker than that observed for electrons. 
The dependence of the electron and gamma-ray fluence on $W$ arises from the fact that 
bremsstrahlung gamma rays emitted by electrons accelerated in the EF region 
tend to spread laterally due to Compton scattering. 
Specifically, when $W$ is small, the Compton scattered gamma rays can easily escape the EF region 
without producing secondary electrons. In contrast, when $W$ is larger, 
the Compton scattered gamma rays have a higher probability of interacting again in the 
EF region and producing additional electrons during their propagation, leading to enhanced 
electron multiplication as suggested by e.g., \citet{dwyer_relativistic_2007}. 
This explains the observed trend that the electron fluence increases with increasing $W$ (Fig.~\ref{fig:electron_photon_L_W_E}).
It should be noted that when $W$ is small compared to $L$, 
particles entering through the lateral
boundaries may contribute non-negligibly since the relative area of the side
surfaces increases. However, as described in the Method section, the
effective path length along $L$ remains shorter for many side-entering particles,
so the dominant contribution to EF-induced multiplication still arises from
particles entering through the upper boundary.

%
%

\subsection{Cumulative energy spectrum expected from secondary cosmic rays}
\subsubsection{Electron spectrum}
\begin{figure}[tbh]
\centering
\includegraphics[scale=0.40]{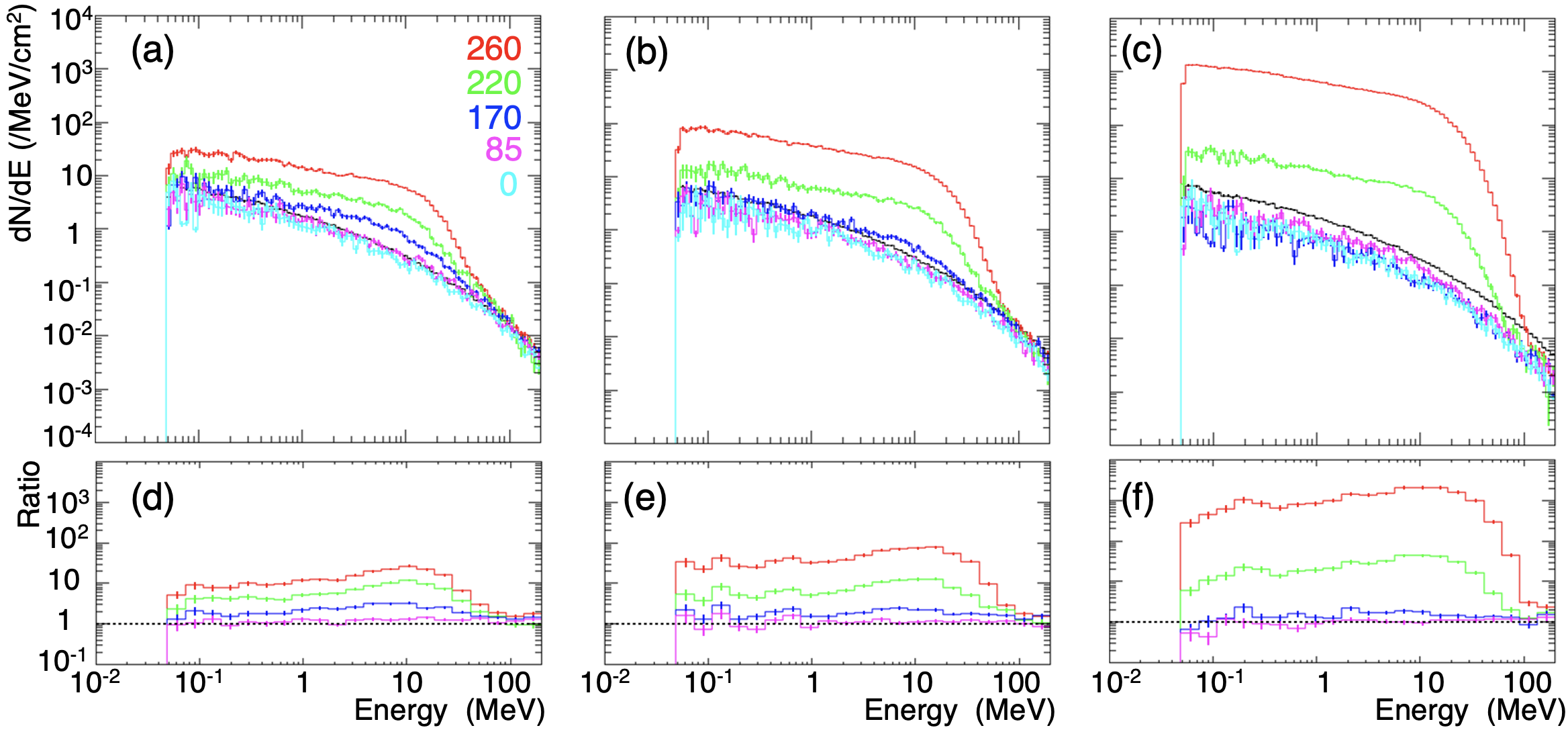}
\caption{Cumulative electron spectra obtained 
at the bottom of the EF region by summing the spectra produced by 
individual incident particles. Examples of the electron energy spectra 
generated by individual injected particles are provided in Supplementary Information, Fig.~S5(a)–(b).
Panels (a), (b), and (c) correspond to $L = 250, 500,$ and 1000~m, respectively. 
Each color line represents a different EF strength
of 0, 85, 170, 220, and 260~$\mathrm{kV\,m^{-1}}$. For comparison, the 
input electron energy spectrum is denoted in black. 
Panels (d), (e), and (f) show the ratios of the electron spectra
for each EF strength to that at 0~$\ef$ (cyan in the upper panels). 
The color meaning in panels (d)$-$(f) is the the same as in panels (a)$-$(c).
}\label{fig:totalespe_ratio}
\end{figure}

To quantitatively determine the constitution of each type of incident particle to electron production,
a cumulative electron spectrum at the bottom of the EF region was obtained via a MC simulation in which 
all secondary cosmic-ray particles were injected (Fig.~\ref{fig:2ndcrspe}).
Figure~\ref{fig:totalespe_ratio} shows the resulting cumulative electron spectra 
for different values of $L$ under several EF strengths.
For EF strengths of 220 and 260 $\ef$,
the absolute electron flux systematically increases as $L$ increases from 250 to 500 and to 1000~m, indicating enhanced electron multiplication in longer EF regions when the field strength exceeds the RREA threshold.
In contrast, 
when the EF strength is 0 $\ef$, electrons injected from just above the EF region 
solely lose energy through ionization as they travel through the region. 
Consequently, for longer propagation distances (i.e., larger $L$), a part of electrons is attenuated, resulting in a spectrum below the background spectrum at the observational level (black) after exiting the EF region. A similar trend is observed when the EF is weak, such as at 85 or 170 $\ef$, where the energy gained from the field is insufficient to compensate for ionization losses.
For clarity, Table~\ref{tab:intflux_ele} lists the ratios of the integrated electron
fluxes for energies above 0.05~MeV at each EF strength relative to that at zero EF.
\begin{table}[tbh]
    \centering
    \begin{tabular}{ccccc}
                & \multicolumn{3}{c}{EF strength ($\ef$)} \\
        $L$ (m) &   260        & 220      &       170       & 85  \\  \hline
         250 &  13.4 $\pm$ 0.3 & 4.73 $\pm$ 0.10 &2.31 $\pm$ 0.05 & 1.23 $\pm$ 0.03 \\
         500 &  44.4 $\pm$ 1.0 & 7.6 $\pm$ 0.2 & 1.92 $\pm$ 0.05  &1.12 $\pm$ 0.03\\
         1000 & 1210 $\pm$ 30  & 25  $\pm$ 5 & 1.52 $\pm$ 0.05  & 1.03 $\pm$ 0.04 \\ \hline
    \end{tabular}
    \caption{Ratios of $>$0.05~MeV electron integral flux at the bottom of the EF region for each EF strength relative to that 
    for 0~$\ef$. $W$ is fixed at 250~m. Each error is a statistical uncertainty derived from MC simulations.}
    \label{tab:intflux_ele}
\end{table}

The ratios in panels (d-f) of Fig.~\ref{fig:totalespe_ratio} indicate that when 
the EF strength exceeds 170~$\ef$, the ratio increases with increasing $L$, as the energy gain from the electric field becomes dominant. In contrast, when the EF strength is below 170~$\ef$, the ratio above 1~MeV decreases with increasing $L$.
When electrons are accelerated in an EF with a strength close to the RREA
threshold, the net energy balance is determined by a competition between
EF energy gain and collisional energy losses. The former increases with both the field strength and the effective acceleration length $L$, and 
the latter, including ionization losses and scattering, are essentially
independent of the EF strength.

For EF strengths at or below the RREA threshold, the energy gain per unit length
is insufficient to compensate for collisional losses.
As $L$ increases, electrons undergo more scatterings, which increases the
probability that they are deflected into an energy range where ionization
losses dominate.
As a result, the ratio above 1~MeV decreases with increasing $L$
[panels (d)--(f) in Fig.~\ref{fig:totalespe_ratio}].
In contrast, when the EF strength exceeds the RREA threshold, the EF
energy gain dominates over collisional losses.
In this regime, a longer $L$ leads to a larger net energy gain and enhances the
development of RREAs, resulting in an increase of the ratio with increasing $L$.

In addition to this energy-balance effect, geometric escape also plays an important role. 
When the width $W$ is small compared with the length $L$ (i.e., $W/L \ll 1$),
electrons can be scattered sideways and escape through the lateral boundaries
before reaching the bottom of the EF region. As a result, many electrons 
effectively experience only a fraction of the EF region, which reduces the 
net energy gain and the electron flux at the bottom.
This side-leakage effect becomes much less significant when $W/L \sim 1$ or
$W/L > 1$.
These effects are directly visible in the two-dimensional ($X$--$Z$) spatial
distributions of electrons; for reference, we show the distribution for
$E = 170~\ef$ in Supplementary Information~4.

The constitution of each type of incident particle was evaluated by calculating the 
electron fluxes above 0.05~MeV from both the cumulative electron spectra and the individual 
spectra for each particle type. Examples of individual spectra are shown in Supplemental information as
Figure 3S(a)-(d) obtained for different $L$ values.
Table~\ref{tab:fraction_elspectr} lists the constitution ratios of these three particle
types in the cumulative electron spectrum. 
Electrons originating from incident gamma rays are found to contribute the most,  
accounting for approximately 60\% of the total electron spectrum. 
This contribution tends to be larger under weaker EF strength and 
appears to be relatively independent of $L$ in the EF region, within statistical uncertainty. 
In contrast, the contribution from incident electrons decreases as the EF strength weakens.
\begin{landscape}
\begin{longtable}{ccccccccccc}
    \caption{Constitution ratios (\%) of incident particles of gamma rays ($\gamma$), 
    electrons ($e^-$), and positron ($e^+$) contributing to the expected electron spectrum at the bottom of the EF region. Statistical uncertainties represent 1$\sigma$ one.}
    \label{tab:fraction_elspectr}\\
                            &\multicolumn{9}{{c}}{$L$ (m)} \\
                            &\multicolumn{3}{c}{250}  &\multicolumn{3}{c}{500}   & \multicolumn{3}{c}{1000}  \\
        EF strength &  $\gamma$   & $e^-$      & $e^+$  & $\gamma$   & $e^-$     & $e^+$ & $\gamma$   & $e^-$      & $e^+$ \\
                $\ef$            & \multicolumn{3}{c}{\%} &  \multicolumn{3}{c}{\%}   & \multicolumn{3}{c}{\%} \\
        260 & $56.2 \pm 1.0$ & $35.2 \pm 0.6$ & $7.6 \pm 0.6$ & $59.1 \pm 0.7$ & $33.1 \pm 0.4$ & $6.8 \pm 0.4$ & $56.5 \pm 0.4$ & $34.9 \pm 0.2$ & $7.7 \pm 0.3$ \\
        220  & $59.1 \pm 1.6$ & $33.9 \pm 0.9$ & $5.7 \pm 0.9$ & $60.7 \pm 1.5$ & $32.1 \pm 0.8$ & $5.2 \pm 0.9$ &$59.2 \pm 1.1$ & $33.5 \pm 0.6$ & $5.3 \pm 0.7$ \\
        170  & $61.0 \pm 2.3$ & $32.0 \pm 1.3$ & $4.9 \pm 1.5$ & $64.3 \pm 2.8$ & $25.2 \pm 1.7$ & $5.7 \pm 1.8$ & $62 \pm 4$ & $18.8 \pm 2.4$ & $6.8 \pm 2.4$ \\
        85 & $63 \pm 3$ & $27.2 \pm 1.9$ & $6.0 \pm 2.0$ & $68 \pm 4$ & $17.6 \pm 2.3$ & $7.6 \pm 2.3$ & $63 \pm 5$ & $12 \pm 3$ & $9 \pm 3$ \\
\end{longtable}
\end{landscape}

\begin{figure}[tbh]
\centering
\includegraphics[scale=0.4]{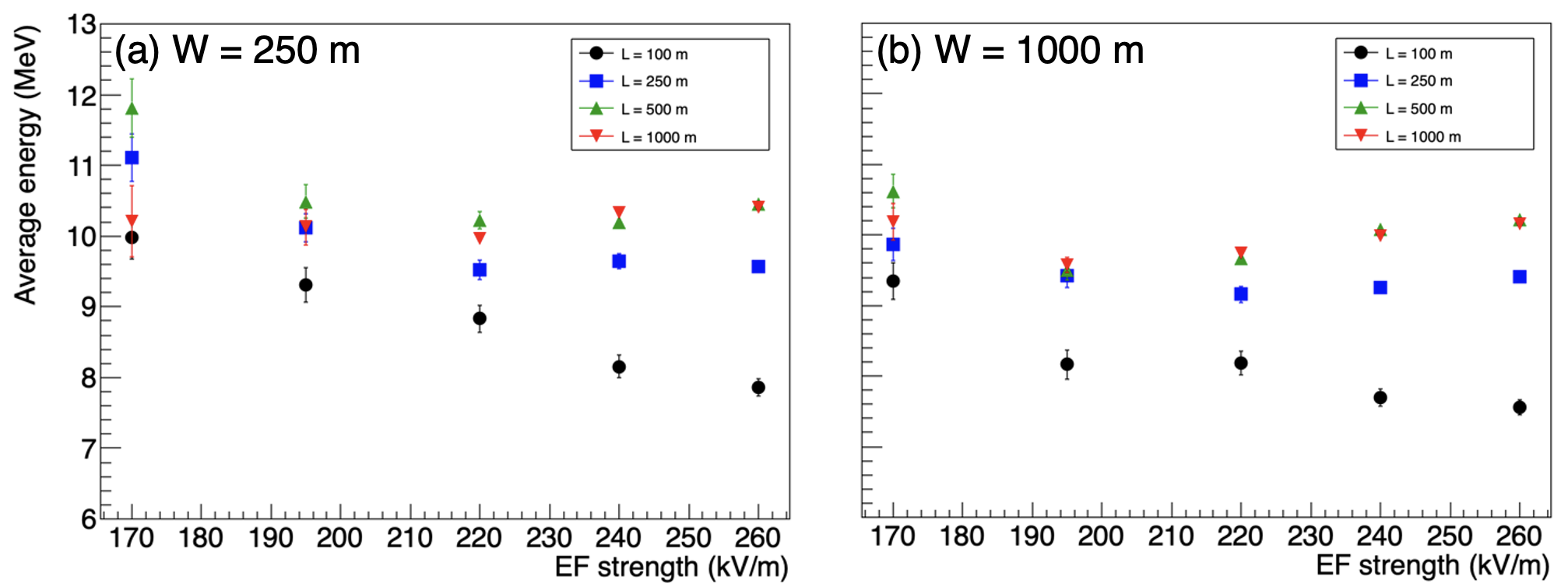}
\caption{Average electron energy obtained by fitting the electron energy spectrum using 
Eq.~\ref{eq:RREAform}. Panels (a) and (b) correspond to $W = 250$  and 1000~m, respectively. 
}\label{fig:aveEne_vs_EF}
\end{figure}

As mentioned earlier, the energy spectra obtained at EF strengths $=$ 260 and 220~$\ef$ (Fig.~\ref{fig:totalespe_ratio}) exhibit shapes similar to  
the RREA energy spectrum reported in previous studies (see, e.g., \cite{dwyer_high-energy_2012} and references therein). 
The RREA energy spectrum is expressed by an exponential function of 
\begin{equation}
    \mathrm{N_0}\exp{(-E_\mathrm{e}/\epsilon_\mathrm{e})}, \label{eq:RREAform}
\end{equation}
where $E_\mathrm{e}$ represents the electron kinetic energy and
$\epsilon_\mathrm{e}$ is the average energy. 
To investigate the dependence on EF strength, the electron spectra obtained for 
various EF strengths were fitted with the above exponential
form over the energy range of 1$-$50~MeV. The resultant values of $\epsilon_\mathrm{e}$ are
plotted as a function of EF strength in Figure~\ref{fig:aveEne_vs_EF}. These results
show that the average electron energy depends on the $L$ value but is relatively insensitive to the $W$ value. 
For $L$ values of 250~m or longer, $\epsilon_\mathrm{e}$ remains about 10--12~MeV 
across the EF strength of 170--260~$\ef$. However, for $L=$ 100~m, the 
$\epsilon_\mathrm{e}$ value decreases with increasing the EF strength.


\subsubsection{Gamma-ray spectrum}
\begin{figure}[tbh]
\centering
\includegraphics[scale=0.40]{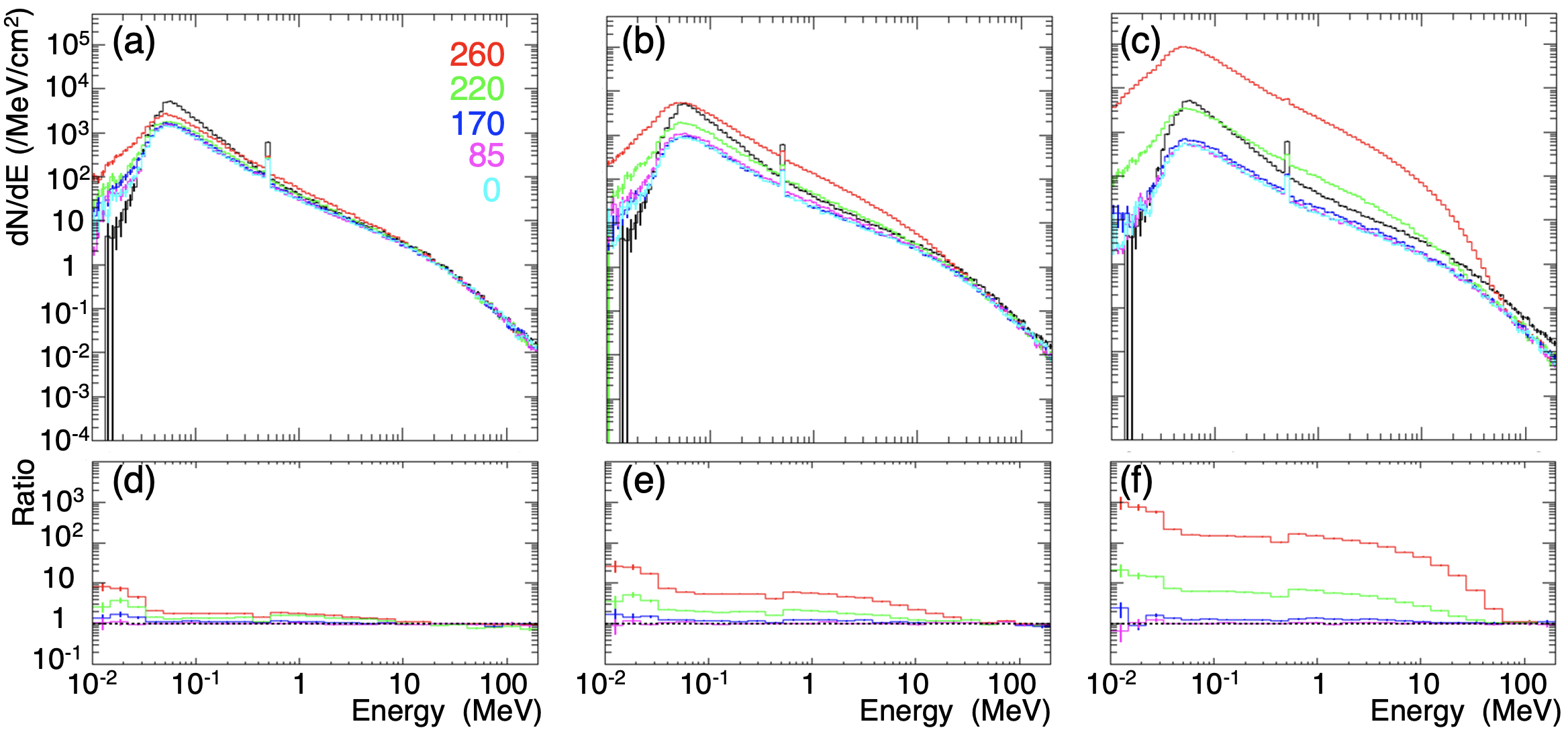}
\caption{The same as Fig.~\ref{fig:totalespe_ratio}, but for cumulative gamma-ray spectra. Examples of the electron energy spectra generated by individual
injected particles are provided in Supplementary Information, Fig.~S5(c)–(d).
}\label{fig:totalgspe_ratio}
\end{figure}

Figure~\ref{fig:totalgspe_ratio} shows the cumulative gamma-ray spectra at the bottom of the EF region. 
The behavior of the gamma-ray cumulative energy spectrum 
is similar to that of the electron energy spectrum (Fig.~\ref{fig:totalespe_ratio}): the spectra
exceed those at 0~$\ef$ (cyan) for longer $L$ values when the EF strength exceeds 
the RREA threshold (170~$\ef$), while they decreases with increasing $L$ below the threshold.

The constitution ratios of incident particles contributing to the gamma-ray 
spectra are 
listed in Table~\ref{tab:fraction_phspectr}. As seen in Fig.~\ref{fig:totalgspe_ratio} and Table~\ref{tab:fraction_phspectr}, gamma rays are the dominant contributors to the cumulative 
gamma-ray spectrum, regardless of the EF strength or $L$ value.
Based on the calculated gamma-ray fluxes (Table~\ref{tab:intflux_ph}), the gamma-ray flux 
increases with increasing $L$. However, this trend is statistically weaker at an EF strength 
of 85~$\ef$. This behavior differs from that of the electron flux (Table~\ref{tab:intflux_ele}), 
which decreases with $L$ when the EF strength is below the RREA threshold.
\begin{landscape}
\begin{longtable}{ccccccccccc}
    \caption{Same as Table~\ref{tab:fraction_elspectr}, but for gamma-ray 
    energy spectra.}
    \label{tab:fraction_phspectr}\\
                            &\multicolumn{9}{{c}}{$L$ (m)} \\
                            &\multicolumn{3}{c}{250}  &\multicolumn{3}{c}{500}   & \multicolumn{3}{c}{1000}  \\
        EF strength &  $\gamma$   & $e^-$      & $e^+$  & $\gamma$   & $e^-$     & $e^+$ & $\gamma$   & $e^-$      & $e^+$ \\
            ($\ef$)           & \multicolumn{3}{c}{\%} &  \multicolumn{3}{c}{\%}   & \multicolumn{3}{c}{\%} \\
        260  & $71.6 \pm 0.6$ & $20.1 \pm 0.4$ & $7.3 \pm 0.4$ & $62.1 \pm 0.5$ & $29.3 \pm 0.3$ & $7.1 \pm 0.3$ & $56.5 \pm 0.4$ & $34.7 \pm 0.2$ & $7.7 \pm 0.2$ \\
        220  & $77.0 \pm 0.6$ & $15.0 \pm 0.4$ & $6.8 \pm 0.4$ & $68.2 \pm 0.6$ & $22.5 \pm 0.4$ & $6.7 \pm 0.4$ & $60.7 \pm 0.6$ & $30.2 \pm 0.3$ & $5.8 \pm 0.4$ \\
        170  & $80.4 \pm 0.7$ & $11.3 \pm 0.5$ & $7.0 \pm 0.5$ & $73.1 \pm 0.8$ & $15.6 \pm 0.5$ & $7.9 \pm 0.5$ & $67.0 \pm 1.0$ & $16.2 \pm 0.6$ & $7.6 \pm 0.6$ \\ 
        85 & $82.3 \pm 0.7$ & $8.6 \pm 0.5$ & $7.8 \pm 0.5$ & $76.1 \pm 0.8$ & $11.2 \pm 0.6$ & $9.0 \pm 0.5$ & $69.0 \pm 1.0$ & $11.4 \pm 0.7$ & $9.0 \pm 0.7$ \\
\end{longtable}
\end{landscape}
\begin{table}[tbh]
    \centering
        \caption{The same as Table~\ref{tab:intflux_ele}, but ratios for fluxes of 
        gamma rays with energies above 0.05~MeV.}
    \begin{tabular}{ccccc}
                & \multicolumn{4}{c}{EF strength ($\ef$)} \\
        $L$ (m) &   260          &   220            & 170           & 85    \\  \hline
         250 &  1.575 $\pm$0.008 & 1.298$\pm$0.008 &1.083 $\pm$ 0.006 &1.009 $\pm$ 0.006 \\
         500 &  4.46 $\pm$ 0.02  & 1.766$\pm$0.012 & 1.152 $\pm$ 0.008  &1.023 $\pm$ 0.007 \\
         1000 & 106.3 $\pm$0.7   & 4.87$\pm$0.04   &1.231 $\pm$ 0.009  & 1.025 $\pm$ 0.009 \\ \hline
    \end{tabular}

    \label{tab:intflux_ph}
\end{table}

\begin{figure}[tbh]
\centering
\includegraphics[scale=0.35]{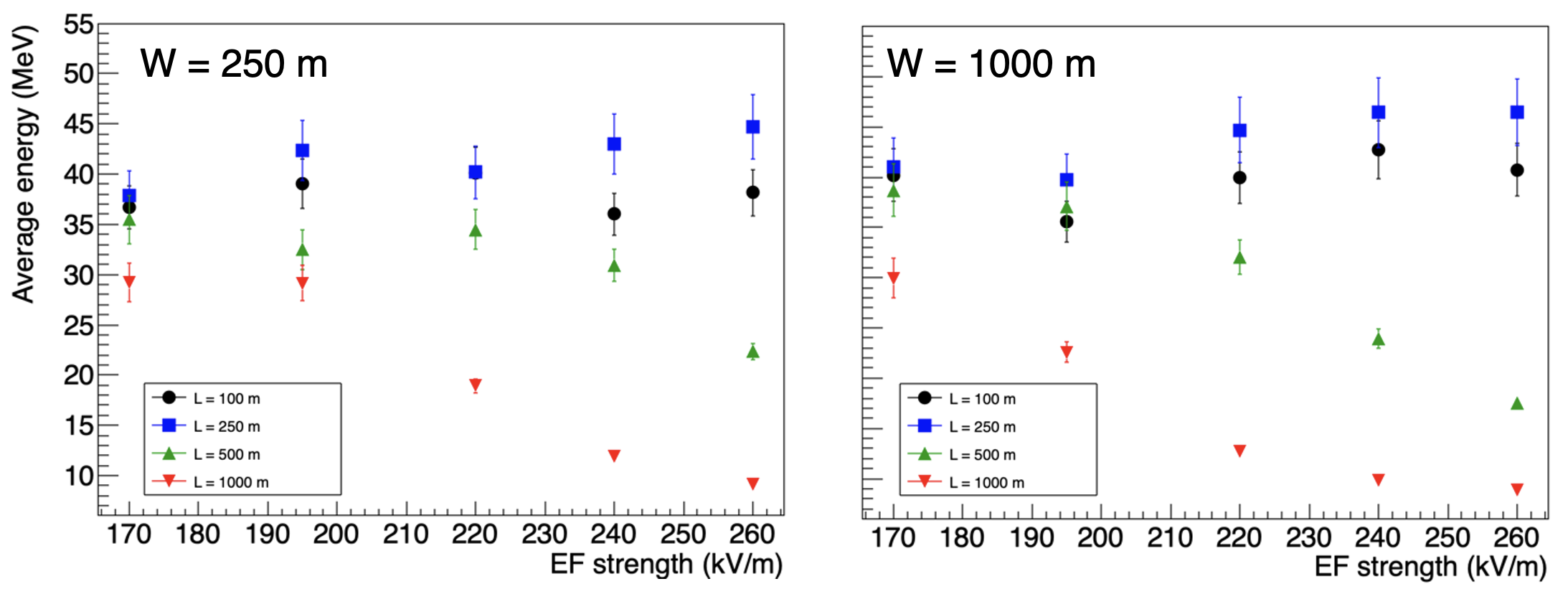}
\includegraphics[scale=0.35]{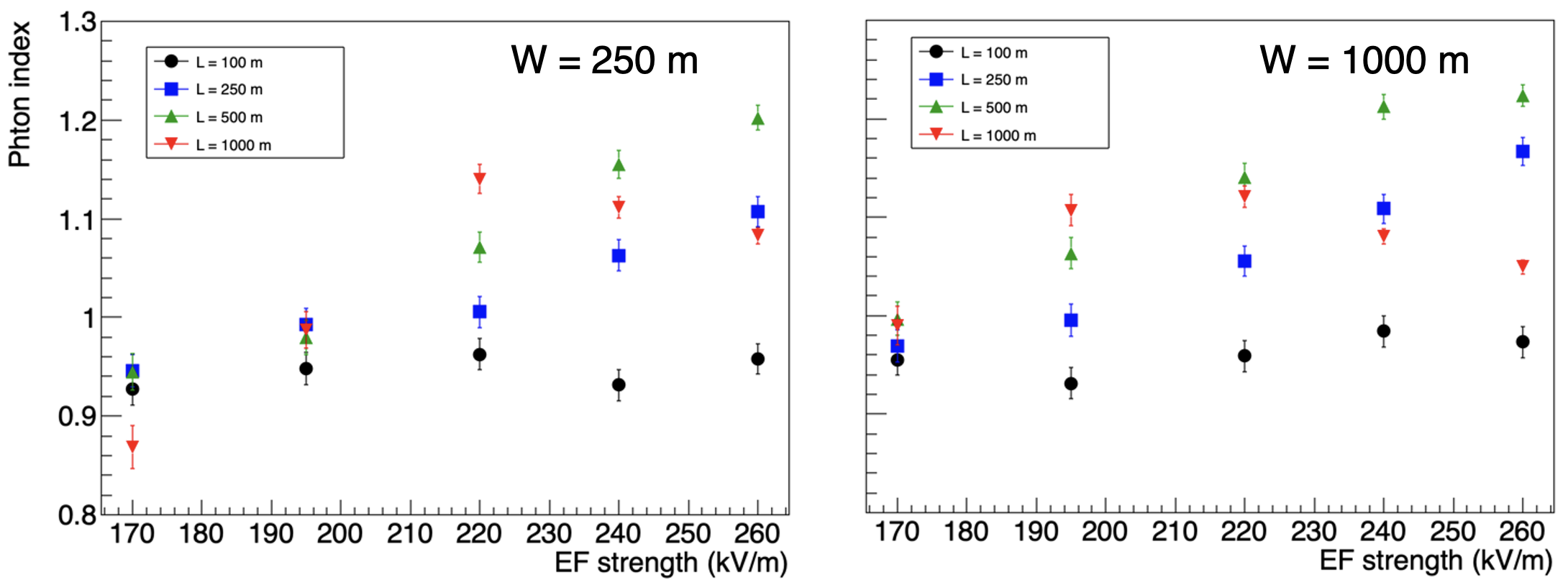}
\caption{Average photon energy (top panels) and power-law index (bottom panels) 
obtained by fitting the gamma-ray
energy spectrum (Eq.~\ref{eq:ph_spe_E260}).
The left panels and right panels correspond to $W=250$~m 
and $W=1000$~m, respectively. 
}\label{fig:gspe_aveEne_pindex_vs_EF}
\end{figure}

The gamma-ray spectra obtained when the EF strength is below the RREA threshold exhibit a shape similar to that of the background gamma-ray spectra expected at the the Yangbajing altitude (Fig.~\ref{fig:totalgspe_ratio}). However, when the EF strength exceeds the RREA threshold, 
the gamma-ray spectra show a different shape. To quantitatively assess this change, 
the gamma-ray spectra at 170, 195, 220, 240, and 260~$\ef$ were fitted in the 1$-$50~MeV energy range 
using the following function:
\begin{equation}
    f(E_\mathrm{p}) = \mathrm{N_0}(E_\mathrm{p}/1\, \mathrm{MeV})^{-\alpha}\exp{(-E_\mathrm{p}/\epsilon_{p})}, 
    \label{eq:ph_spe_E260}
\end{equation}
where $E_\mathrm{p}$ is the gamma-ray energy in MeV, 
$\mathrm{N_0}$ is a normalization constant, 
$\alpha$ is the photon index, and 
$\epsilon_{p}$ represents the average photon energy. 
This spectral form is based on the assumption that the gamma rays are
produced by electrons undergoing RREA as expressed by Eq.(\ref{eq:RREAform}).
The resulting fit parameters, $\epsilon_\mathrm{p}$ and $\alpha$, are shown in Figure~\ref{fig:gspe_aveEne_pindex_vs_EF}. 
As indicated, when $L$ exceeds 500~m, the $\epsilon_\mathrm{p}$ tends to decreases 
with increasing EF strength.
It is notable that the $\epsilon_\mathrm{p}$ at EF strength of 260~$\ef$ is approximately 
equivalent to that of electrons ($\sim$9~MeV).
Conversely, when $L$ is reduced to 250~m or 100~m, 
the $\epsilon_\mathrm{p}$ reaches approximately 40~MeV. 
The average photon energy appears relatively insensitive to changes in $W$. 
The photon index $\alpha$ tends to increase with both increasing $L$ and increasing 
the EF strength. 

\subsection{Dependence on cloud base heights}
\begin{figure}[tbh]
\includegraphics[scale=0.40]{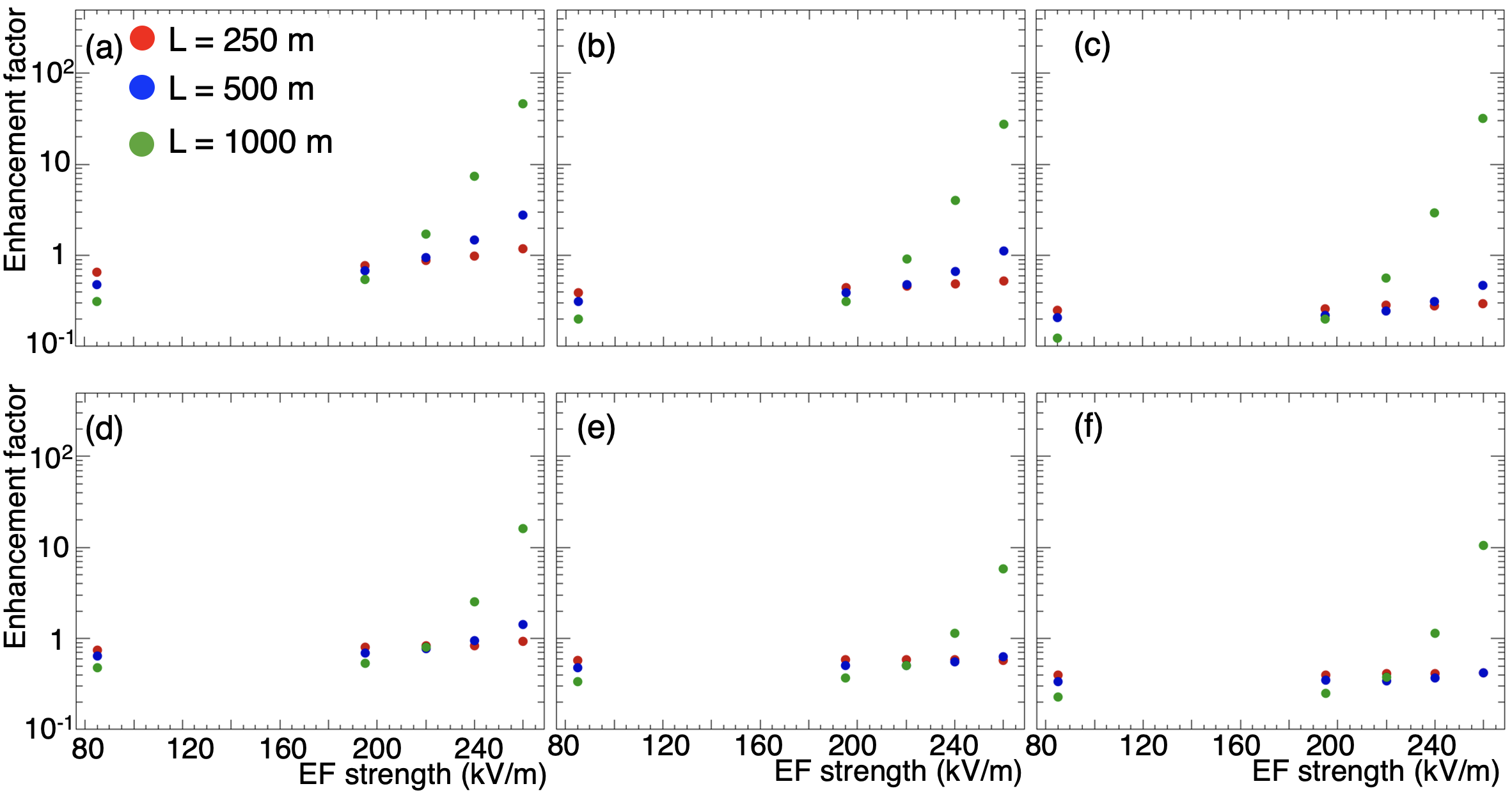}
\caption{Enhancement factors as a function of EF strength ($\ef$). The top panels 
show enhancement factors for 0.01$-$10~MeV energy range, while the bottom panels denote
those for $>$10 MeV. From left to right, the panels correspond to $H=50, 500,$ and 1000~m.
}\label{fig:EF_vs_Efactor}
\end{figure}

A cloud base height (denoted as $H$ in this paper) refers to the vertical 
distance from the ground surface to a thundercloud base. This parameter affects the particle flux in long-duration bursts that can be detected at the observation level. Figure~\ref{fig:EF_vs_Efactor} shows enhancement factor, defined as a ratio of the total arriving particle flux (including gamma rays, electrons, positrons, and muons) from a EF region to the background flux at the Yangbajing altitude. The background flux refers to the flux of secondary particles obtained from PHITS simulations without applying an electric field, assuming both $H$ and $L$ are set to 1~m.
It is found that the enhancement factor increases as the EF strength becomes stronger and/or the length of the EF region ($L$) becomes longer. Particularly, 
for $L= 1000$~m, the enhancement factors in all conditions tend to increase steeply as the EF strength increases. 
These enhancement factors represent the particle fluxes arriving just above the detector location at the Yangbajing altitude,without accounting for 
effective areas of detectors. 

\begin{figure}[tbh]
\centering
\includegraphics[scale=0.30]{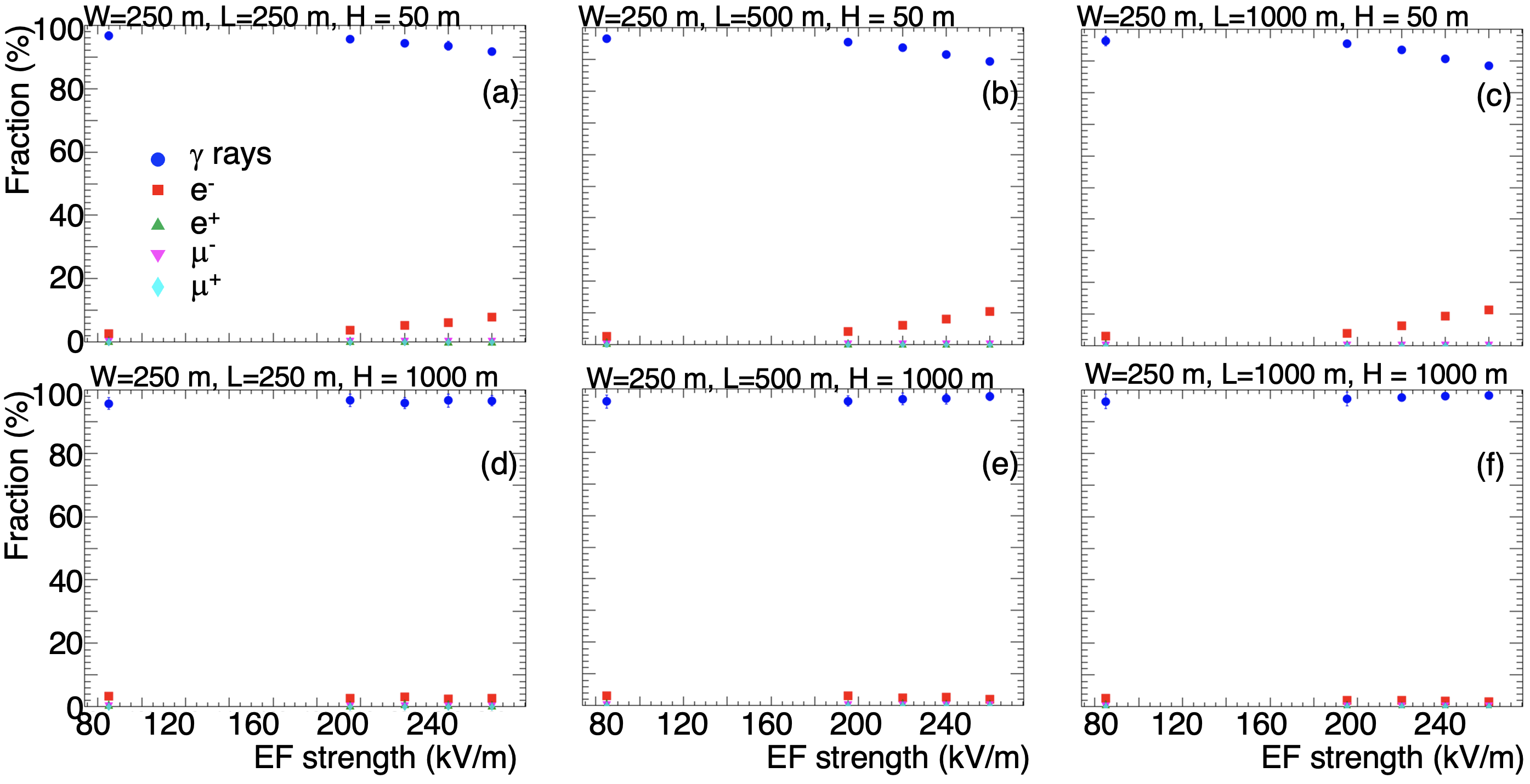}
\caption{Fraction (in percent) of each particle type in the 0.01$-$10~MeV energy 
range arriving at the ground surface, shown as a function of EF strength ($\ef$). 
All panels assume a fixed $W=250$ m. The top row (panels a$-$c) represents
results for $H=50$ m, while the bottom row (panels d$-$f) shows results for $H=1000$~m. 
From left to right, the panels correspond to $L=250, 500$, and 1000~m. Circle (blue) : gamma rays, Square (red) : electrons, Triangle (green) : positrons, Inverted triangle (magenta) : negative muons, Diamond (cyan) : positive muons
}\label{fig:lowE_EachFrction_ground}
\end{figure}

\begin{figure}[tbh]
\centering
\includegraphics[scale=0.30]{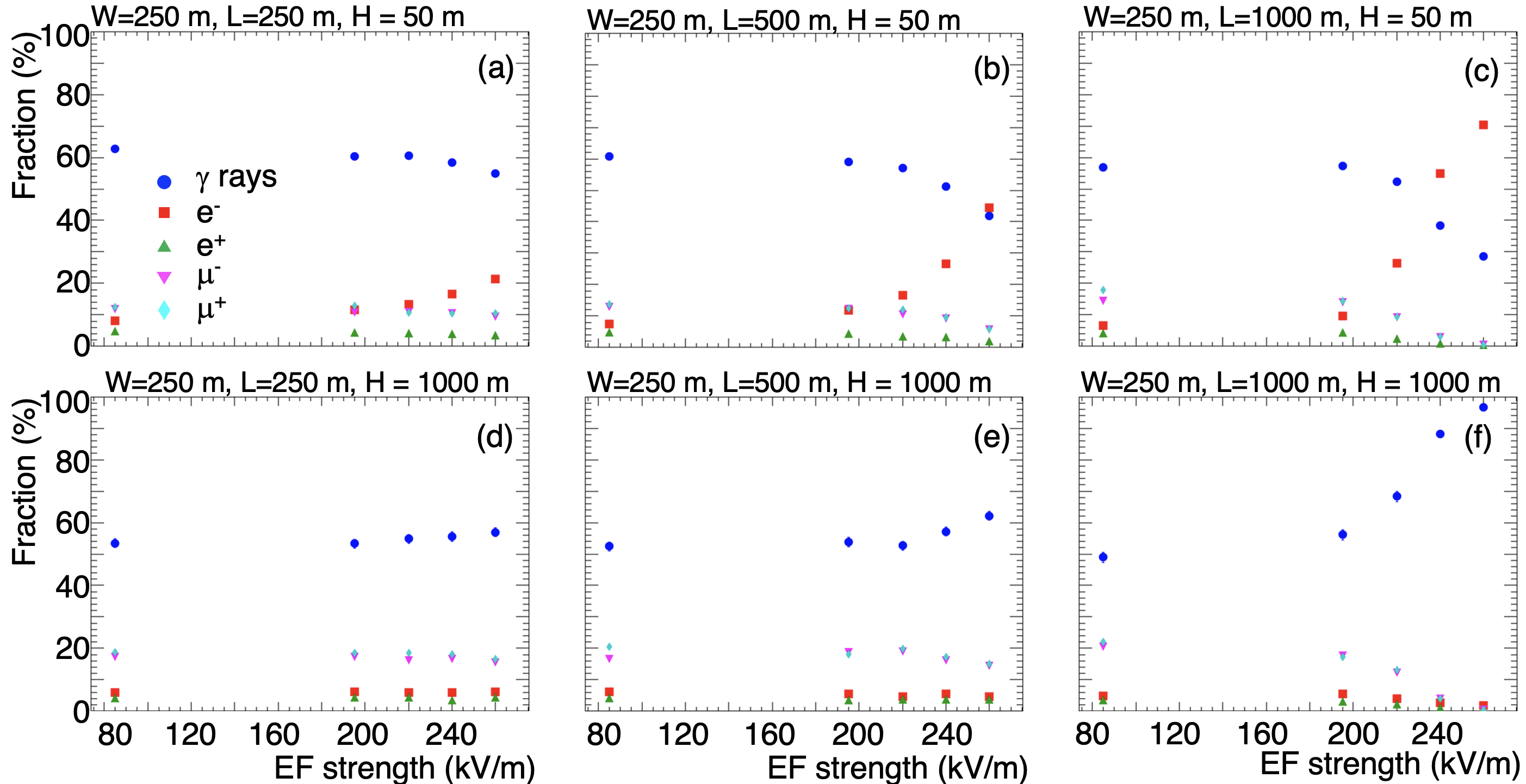}
\caption{The same as Fig.~\ref{fig:lowE_EachFrction_ground}, but for $>$10~MeV 
particle types.}\label{fig:highE_EachFrction_ground}
\end{figure}

It is important to understand which types of particles are likely to arrive just above detectors. 
Figures~\ref{fig:lowE_EachFrction_ground} and ~\ref{fig:highE_EachFrction_ground} present the relative contributions of gamma rays, electrons, positrons, negative muons, and positive muons 
that reach the ground surface after traveling in an EF region.
From Fig.~\ref{fig:lowE_EachFrction_ground}, it can be seen that not depending on EF conditions, fractions of gamma rays account for approximately 90\% or more of the arriving particles in the 0.01$-$10~MeV energy range. In this energy range, 
short range of electrons ($<$6 m) prevents most of electrons from reaching detectors. 
Also, since the muon contribution is inherently small in this energy range (Fig.~\ref{fig:2ndcrspe}), the muon component affected by the electric field 
remains negligible. 

As shown in Fig.~\ref{fig:highE_EachFrction_ground}, the fractions of gamma rays and electrons with energies above 10~MeV are largely affected by the EF conditions. 
When $H$ is 50~m (upper panel of Fig.~\ref{fig:highE_EachFrction_ground}), 
the gamma-ray fraction decreases, while electron fraction increases with increasing EF strength.
Notably, when the $L$ exceeds 500~m, electrons become the dominant component 
under strong EF strengths.
These results imply that electrons become the primary particles observed when a thundercloud is located very close to detectors.
In contrast, when the $H$ is 1000~m (lower part of Fig.~\ref{fig:highE_EachFrction_ground}), 
gamma rays constitute more than half of the particles arriving at the ground surface, 
and their fraction increases with EF strength. 
Muons constitute the next largest contribution, 
likely due to their much longer penetration range compared to 
electrons and positrons, especially at larger $H$ values. 
The muons reaching the ground surface are not newly 
produced in an EF region, but are incident particles that lost energy mainly via ionization loss while traversing the EF region and then propagate in the atmosphere.

\section{Discussions}
\subsection{Gamma-ray contribution}
The present study indicates that the gamma-ray component of steady-state secondary cosmic rays 
makes a substantial contribution to the production of electrons and gamma rays 
that escape from an EF region (Tables~\ref{tab:fraction_elspectr} and \ref{tab:fraction_phspectr}). 
At both high altitudes and sea level,
the flux of gamma rays in the 0.01~MeV to 1~GeV energy range exceeds 
that at of electrons and positrons. For example, according to EXPACS~\cite{EXPACS}, 
the integrated fluxes (cm$^{-2}$~s$^{-1}$) from 0.01~MeV to 1~GeV are 1.3 for gamma rays, 
$3.7\times10^{-2}$ for electrons, and $1.6\times10^{-2}$ for positrons at an altitude of 4.3~ km (Yangbajing). At sea level in Japan, the corresponding fluxes are 
$1.0\times10^{-1}$ for gamma rays, $3.0\times10^{-3}$ for electrons, and $1.4\times10^{-3}$ for positrons. 
Given these fluxes, the present results indicate that 
gamma rays play a dominant role in particle enhancement by supplying seed electrons via electromagnetic interactions inside the EF region.

This mechanism operates under steady-state conditions and is fundamentally 
different from particle production in an extensive air shower (EAS).
The EAS is known to generate transient gamma-ray cascades initiated by individual primary cosmic rays. 
On the other hand, the secondary cosmic rays considered here form a persistent 
particle background that continuously interacts with thundercloud electric fields
over much longer timescales. 
Therefore, the dominance of gamma rays discussed here reflects their importance
in the steady-state secondary cosmic-ray environment, where both their high ambient
flux and their ability to generate electrons via electromagnetic interactions make
them the primary suppliers of seed electrons for EF-induced acceleration, rather
than their transient abundance in extensive air showers.
This result may have implications for modeling electron acceleration inside thundercloud electric fields. 
Since the seed population affects electron avalanche development, incorporating realistic secondary gamma-ray fluxes 
could help improve the physical consistency of simulations. 
A quantitative comparison of the relative contributions of different secondary cosmic-ray components to seed-electron production remains an important subject 
for future investigation.
\subsection{Average electron energy}
According to previous studies, the average energy of electrons ($\epsilon_\mathrm{e}$) produced by RREA is approximately 7~MeV~\cite[references therein]{dwyer_fundamental_2003, dwyer_high-energy_2012}. 
In contrast, the present study finds the $\epsilon_\mathrm{e}$ value to be 9–10~MeV (Fig.~\ref{fig:aveEne_vs_EF}). 
\citet{sarria_2018_GEANT4sim} utilized GEANT4 simulations to estimate the $\epsilon_\mathrm{e}$, 
reporting $\epsilon_\mathrm{e}$ of 7-9~MeV for EF strength ranging 
from 0.5 to 3~MV~m$^{-1}$. When employing the 
GEANT4 Option 1 model (standard electromagnetic interactions), their results yielded values of 
9–12~MeV, which are comparable to those obtained in this study.
Similarly, \citet{skeltved_modeling_2014} showed that $\epsilon_\mathrm{e}$ reaches 
approximately 10~MeV when using a different GEANT4 model (Low- and High-Energy Parameterization, LHEP). These differences in $\epsilon_\mathrm{e}$ values may therefore be attributable to the 
different interaction models assumed in PHITS, GEANT4 and other simulations.

In the present study, $\epsilon_\mathrm{e}$ approaches 7~MeV when $L = 100$~m,
indicating that $L$ influences on the resulting $\epsilon_\mathrm{e}$.
Previous studies, such as \citet{dwyer_fundamental_2003} and \citet{skeltved_modeling_2014},
employed $L \sim 200$m at 1~atm.
To examine the origin of the higher $\epsilon_\mathrm{e}$ values obtained here, 
we conducted additional simulations by vertically injecting monoenergetic 1~MeV electrons into EF regions, with varying $L$.
The resulting $\epsilon_\mathrm{e}$ at 260~$\ef$ were 3.6, 6.6, 10.4, and 10.3~MeV for $L=100$, 250, 500, and 1000~m, respectively. 
These results suggest that $L$ plays a key role in determining $\epsilon_\mathrm{e}$. 
\subsection{Maximum gamma-ray energy}\label{sec:max_photon_E}
In the long-duration bursts observed in Yangbajing, 
gamma rays with energies 40~MeV or higher have 
been detected by a solar neutron detector~\cite{tsuchiya_observation_2012}. To investigate the structure and strength of 
the EF capable of producing such high-energy gamma rays, we analyzed the ratios shown in Fig.~\ref{fig:totalgspe_ratio}. We identified the energy range above 1~MeV where the ratio exceeds 1, and defined its upper bound as the maximum gamma-ray energy, $K_\mathrm{max}$, expected from an observation. 

Table~\ref{tab:Kmax_EF} summarizes the results, and shows that 
$K_\mathrm{max}$ increases with $L$. This is consistent with the fact that electrons gain 
more energy while traveling in longer distances within the EF region
until energy gain and loss reach equilibrium. The $K_\mathrm{max}$ value appears to be relatively 
insensitive to $H$, since similar values are obtained for both $H =50$~m and 1000~m.
Table~\ref{tab:Kmax_EF} also indicates that EF conditions 
capable of generating gamma rays 
with energies above 40~MeV require $L \ge 500$~m and an EF strength of $\ge$ 220~$\ef$. 
%
\begin{table}[thbp]
  \centering
  \caption{Maximum gamma-ray energy (MeV) expected under several EF conditions.}
  \begin{tabular}{cccccccc}
    \multicolumn{4}{c}{EF condition} & \multicolumn{4}{c}{$K_\mathrm{max}$ (MeV)} \\
    $W$ (m) & $L$ (m) & $H$ (m) & & 260 $\ef$ & 220 $\ef$ & 170 $\ef$ & 85 $\ef$ \\
    \hline
    250 & 250  & 50   & & 29  & 4.0  & 4.0  & 1.1 \\
    250 & 500  & 50   & & 43 & 39 & 7.3 & 1.3 \\
    250 & 1000 & 50   & & 95  & 43 & 26  & 2.7 \\
    250 & 250  & 1000 & & 15  & 15  & 1.8 & 1.8 \\
    250 & 500  & 1000 & & 43 & 26  & 3.3 & 1.3 \\
    250 & 1000 & 1000 & & 95  & 58  & 22  & 1.3 \\
    1000 & 250 & 50   & & 16  & 12  & 2.4  & 1.1 \\
    1000 & 500 & 50   & & 71  & 43 & 6.0  & 1.5 \\
    1000 & 1000 & 50  & & 78  & 71  & 26  & 1.5 \\
    \hline
  \end{tabular}
  \label{tab:Kmax_EF}
\end{table}
\subsection{Comparison with the past observations at Yangbajing}
The Yangbajing neutron monitor (NM) detects many prolonged gamma-ray emissions~\cite{tsuchiya_tibet_2024}.
Rather than attempting a direct and detailed reproduction of individual NM observations, 
which is beyond the scope of the present study, we use these observations to perform
a consistency check on whether the modification of secondary cosmic rays obtained for various
EF conditions can plausibly account for the magnitude of
long-duration bursts actually observed at Yangbajing. To this end, we conduct an
order-of-magnitude estimate of the particle flux enhancement corresponding to the
simulation results shown in Fig.~\ref{fig:highE_EachFrction_ground}.

Using the count increases and event durations shown in Table~1 of \citet{tsuchiya_tibet_2024}, 
along with the NM area of 32~m$^2$, the observed gamma-ray flux is estimated to be approximately in the range
10$^{-5}$ to 10$^{-4}$~cm$^{-2}$~s$^{-1}$. The NM detection efficiency for gamma rays 
in the 10–200~MeV energy range is on average about $10^{-3}$~\cite{tsuchiya_observation_2012}. 
When the observed flux is divided by this detection efficiency, a gamma-ray flux arriving just above the NM of 
0.01 to 0.1~cm$^{-2}$~s$^{-1}$. 
Under the present simulation conditions, 
the calculated fraction of gamma rays among all thundercloud-related particles ranges from 30\% to 95\% (Fig.~\ref{fig:highE_EachFrction_ground}). Accordingly, the total flux of such particles just above the NM 
is estimated to lie in the range of 0.01 to 0.3~cm$^{-2}$~s$^{-1}$.
For comparison, the background flux of $>$10~MeV particles is calculated to be
0.21~cm$^{-2}$s$^{-1}$ by integrating the EXPACS data for each particle type (Fig.~\ref{fig:2ndcrspe}). 
Therefore, the enhancement factor, obtained by taking the ratio of 0.01/0.21 and 0.3/0.21, 
is estimated to range from 0.05 to 1.4. 

This comparison shows that the EF parameter range explored in this study
produces particle enhancement levels comparable to those inferred
from NM observations. Although this approach
does not allow us to place stringent constraints on the exact EF strength or
geometry, it demonstrates that the present simulations cover physically
realistic conditions relevant to observed long-duration bursts.

\subsection{Electrons arriving at the ground surface}
It has been argued that during the production of long-duration bursts, not only gamma rays but also accelerated electrons 
may be directly detected. According to \citet{chilingarian_comment_2023}, 
such electron detection become possible when the CBH (or $H$ in this work) is particularly low, between 25 and 150~m. 
Their CORSIKA~\cite{CORSIKA} simulations show that when $H= 25$~m, the ratio of $>$10~MeV electrons to $>$10~MeV gamma rays just above their detectors amounts to be 1.44, whereas it drops to 0.011 at $H=200$~m. 
In addition, their observations~\cite{chilingarian_comment_2023} indicates that 
electron-to-gamma-ray ratios range from 0.11 to 0.26. Comparable values can be derived from the fraction values in Fig.~\ref{fig:highE_EachFrction_ground}, by computing a ratio of 
an electron faction (red points) to a gamma-ray fraction (blue points). The ratio at $H =$ 50~m (upper panels in Fig.~\ref{fig:highE_EachFrction_ground}) ranges from 0.1 to 2.3, 
while it falls to between 0.02 and 0.1 at $H = 1000$~m (bottom panels in Fig.~\ref{fig:highE_EachFrction_ground}). The derived ratio of 0.1-2.3 for $H = 50$~m, is consistent with the values 
reported by \citet{chilingarian_comment_2023}, despite the difference in observational altitudes (3200~m of Mt. Aragtz versus 4300~m of Yangbajing).

\citet{williams_ConditionsEnergeticElectrons_2022} also discussed the detection of accelerated electrons using
GEANT4 simulations, estimating an electron-to-gamma-ray ratio of 0.40 at $H = 50$~m. They further 
emphasized the significance of nocturnal events, pointing out that CBH tends to be lower at night 
due to surface cooling, which brings the temperature closer to the dew point. 
This suggests that accelerated electrons related to long-duration bursts are more likely to be detected in nighttime. 
Indeed, observations at Yangbajing~\cite{tsuchiya_tibet_2024} and along the coastal area 
of the Japan Sea~\cite{wada_catalog_2021} indicate that long-duration bursts frequently occur
during local nighttime. A detailed investigation of such nocturnal events,
as suggested by \citet{williams_ConditionsEnergeticElectrons_2022},
may therefore provide further insights into the detection of accelerated electrons.

\subsection{Comparisons with other simulations}
In this study, we used PHITS to investigate the production of electrons and gamma
rays in atmospheric electric fields. Related studies have been conducted with other simulation codes. 
In particular, \citet{Zhou_2016_AROG_YBJ} examined EF-induced variations of electrons and positrons in air 
showers under EFs at the Yangbajing site (4300 m a.s.l.), using the CORSIKA simulation~\cite{CORSIKA}, 
which is widely employed in cosmic-ray physics. 
To enable a direct comparison and understand systematic uncertainty associated with different codes, 
we adopted the same EF configuration as \citet{Zhou_2016_AROG_YBJ}.

It should be noted that the treatment of incident particles differs
between the two studies. In \citet{Zhou_2016_AROG_YBJ}, secondary particles
entering the EF region are generated by vertically injecting GeV-order primary
protons at the top of the atmosphere and tracking the resulting air showers
down to 2.0~km above Yangbajing. In contrast, in the present study, secondary
cosmic rays are directly injected from the top of the EF region, with an
angular distribution corresponding to zenith angles up to 60$^\circ$.
Therefore, the physical situation treated here is 
fundamentally different from that in \citet{Zhou_2016_AROG_YBJ} and 
corresponds to the steady-state modification of ambient secondary cosmic rays by thundercloud
electric fields, rather than the transient transport of a single air shower through an EF.

\begin{figure}[tbh]
\centering
\includegraphics[scale=0.35]{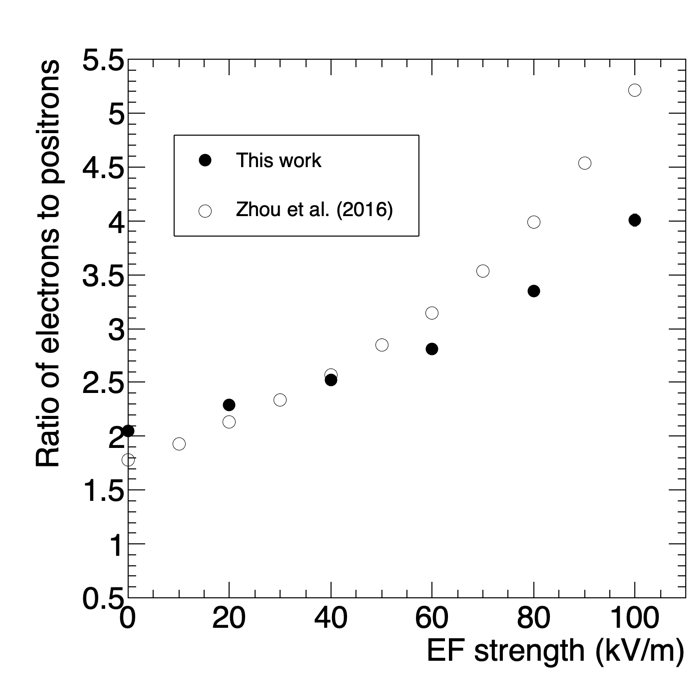}
\caption{Ratio of electron flux to positron flux above 0.1~MeV 
at Yangbajing 
level (altitude of 4300 m). The horizontal axis shows the EF strength, and the vertical 
axis represents the flux ratio. Open  circles indicate the results of \citet{Zhou_2016_AROG_YBJ}, and filled circles show the PHITS results (this work).
}\label{fig:comp_NeNp_Zhou2016}
\end{figure}

Figure~\ref{fig:comp_NeNp_Zhou2016} shows the ratio of electron flux to positron flux at energies above 0.1~MeV at the ground level (4300~m a.s.l.), calculated using PHITS under an EF condition with $L$ = 2000~m (4300--6300~m in the vertical direction) and $H$ = 0~m. For comparison, the corresponding results
reproduced from Fig.~6 of \citet{Zhou_2016_AROG_YBJ} are also shown. 
Their CORSIKA results correspond to vertically incident 100~GeV primary protons. 
For consistency, we use the EF polarity that accelerates electrons downward, as shown in their paper. 
The two simulations agree reasonably well for moderate EF strengths.
The discrepancy increases with EF strength and reaches approximately 30\% at 100~$\ef$.

An important implication of this comparison is that, despite the different
treatments of incident particles, the EF-induced variations in electron and
positron fluxes at ground level differ by less than approximately 30\% over
the EF range from 0 to 100~$\ef$.
This general agreement is not trivial and indicates that the amplification process inside the EF region is
largely insensitive to how the particles were originally produced and is
instead governed mainly by the local secondary cosmic-ray spectrum entering
the EF region. At the same time, the differences between PHITS and CORSIKA highlight 
the level of systematic uncertainty
that should be considered when comparing simulation results of electron
acceleration in thundercloud electric fields.
A more detailed comparison among PHITS, CORSIKA, and GEANT4 is provided 
in the Supplementary Information (Figure~S6).

\section{Summary}
Using Monte Carlo simulations by PHITS, we simulated the behavior of secondary cosmic rays within an EF region to investigate how they produce electrons and gamma rays detectable at an altitude of 4.3 km. 
Unlike most previous simulations, which typically assume vertically injected monochromatic 
1~MeV electrons, we adopted a more realistic approach. Specifically, the energy spectra of secondary cosmic rays derived from EXPACS~\cite{EXPACS} were used to simulate the initial 
energy distribution of particles entering the electric field region. 
In addition, the zenith angle distributions of the incident particles 
were set to range from 0$\tcdegree$ to 60$\tcdegree$, reflecting the actual angular spread of secondary cosmic rays.

This study yields the following main results. 
Among the components of secondary cosmic rays, gamma rays play a dominant role in supplying high-energy electrons inside the EF region, 
leading to the observed particle enhancement at an altitude of 4.3~km. This dominance arises from a combination of their ambient flux under steady-state atmospheric 
conditions and their efficiency in producing energetic electrons through electromagnetic interactions within the EF region.

The average energy of electrons produced from the RREA process was found to be slightly higher than the conventional value of around 7 MeV, reaching approximately 9--10 MeV. 
This difference is due to the fact that the conventional values were derived 
from MC simulations using monoenergetic 1 MeV electrons, whereas 
the present study considered incident electrons with higher energies.

A comparison was also made between the simulation results and measurements obtained by neutron monitors and a solar neutron telescope, which are both located at Yangbajing, Tibet. The analysis showed that the EF conditions capable of producing high-energy gamma rays, as observed in Tibet, correspond to EF strengths of $\ge$~220~$\ef$ and EF region lengths of at least 500~m. 




\section*{Acknowledgments}
This research was conducted with the supercomputer HPE SGI8600 in the Japan Atomic Energy Agency (JAEA). 
I would like to thank the PHITS developers, including Dr. T. Sato and 
Dr. T. Ogawa (both at JAEA), for their valuable support regarding its usage. 
I am also grateful to Prof. E. Williams and Dr. Y. Wada for their stimulate discussions 
on nighttime observations. 
\newpage

\bibliographystyle{elsarticle-num-names} 
\bibliography{ref202408}





%
\end{document}